\documentstyle[emulateapj,psfig]{article}
\def\lsim{\lower.5ex\hbox{$\; \buildrel < \over \sim \;$}}
\def\gsim{\lower.5ex\hbox{$\; \buildrel > \over \sim \;$}}

\def\lsim{\lower.5ex\hbox{$\; \buildrel < \over \sim \;$}}
\def\gsim{\lower.5ex\hbox{$\; \buildrel > \over \sim \;$}}

\received{Accepted for publication in {\it Astrophysical Journal Letter}}
\begin{document}

\title{Origin and interpretation of  kilohertz QPOs from strange stars
in X-ray binary system: theoretical hydrodynamical description}

\author{Banibrata Mukhopadhyay$^1$, Subharthi Ray$^2$, Jishnu Dey$^3$,
Mira Dey$^4$}  
\affil
{\small 1. Inter-University  Centre for Astronomy and Astrophysics,  
Ganeshkhind, Pune, India\\
2. Instituto de Fisica, Universidade  Federal Fluminense,  
Niter\'oi, RJ, Brazil; FAPERJ Fellow\\
3. Abdus Salam ICTP, Trieste, Italy; on leave
from Maulana Azad College, Kolkata, India\\
4. Abdus Salam ICTP, Trieste, Italy; on leave from Presidency College, Kolkata,
India\\}

\begin{center} Accepted for publication in {\it Astrophysical Journal Letter} \end{center}

\begin{abstract}
We  model and  interpret the  Kilohertz QPOs  from  the
hydrodynamical description of accretion disk around  a rapidly rotating compact
strange star. The higher QPO frequency  is  described by  the
viscous  effects  of accretion  disk leading to  shocks, while
the lower  one is taken to  be the Keplerian motion  of  the
accreting  matter.  Comparing our  results  with  the
observations  for   two  of  the  fastest   rotating  compact
stellar candidates namely,  4U~1636$-$53 and KS~1731$-$260, we
find that they match to a very good  approximation, thus
interpreting them as strange stars.
\end{abstract}

\keywords {accretion, accretion disks --- hydrodynamics --- shock
waves --- stars: neutron --- stars: individual (4U 1636-53, KS 1731-260) 
--- X-rays: binaries  }

\section{Introduction}

In  X-ray binaries,  matter  is transferred  from  a normal  star to  a
compact star. As the characteristic velocities near the compact object
are of  order $\rm (GM/R)^{1/2}\sim 0.5c$, the (dynamical) time scale
for the  motion of matter through  the emitting region,  is short. So,
the significance of millisecond  X-ray variability from X-ray binaries
is clear  from the  phenomenon that milliseconds  is the  natural time
scale  of the  accretion process  in the  X-ray emitting  regions, and
hence strong X-ray variability on such time scales is certainly caused
by  the motion  of matter  in these  regions. Orbital  motion, stellar
spin, disk  and the  stellar oscillations are  all expected to  happen on
these time scales.

In recent years, kilohertz quasi-periodic oscillation (kHz QPOs) peaks,
with a  certain width, have  been found in  the power spectra  of more
than 20  low-mass binaries (LMXBs). In  most of the sources, the  kHz QPOs are
found  in pairs  and at  least  one of  the peaks  should reflect  the
orbital motion  of matter  outside, but not  too far from  the compact
star, in near-Keplerian orbits.

Another high frequency phenomenon, namely nearly coherent oscillations
which slightly  drift in  frequency, was detected  during some  type I
X-ray bursts. These so-called burst oscillations have been detected in
the power spectra of some  ten sources and their X-ray flux modulation
is consistent with being due to  the changing aspect of a drifting hot
spot  on  the surface  of  the  compact  star. Therefore, these  burst
oscillations are thought to reflect the spin frequency of the star.

In a  very important work, Jonker et al. (2002, hereinafter JMK) report on high 
frequency QPO phenomena that have been observed in the LMXBs, particularly 4U 1636$-$53;
two kHz QPOs (Zhang et al. 1996a; Wijnands et al. 1997)  as well
as a sideband to the lower of the  two kHz QPOs (Jonker et al. 2000)
and burst oscillations with frequency
$\nu_b  = 581$ Hz (Zhang et al. 1996b;  Strohmayer  et al. 1998).
Miller (1999) presented evidence that  these oscillations are  in
fact the second harmonic of  the spin  of the  star $\sim 290.5$ Hz.
However, using another  data set  these  findings  were not
confirmed (Strohmayer 2001).

Since the frequency difference between the two kHz QPO peaks (the peak
separation $\Delta \nu$) is nearly equal to half the burst oscillation
frequency,  a beat  frequency  model  was proposed  for  the kHz  QPOs
(Strohmayer et al. 1996). In such  a model, the higher frequency QPO is
attributed,  as in  most other  models,  to the  orbital frequency  of
chunks of plasma  at a special radius near the  compact star while the
lower QPO is due to beat between the orbital frequency and the stellar
spin  frequency. A  specific  model incorporating  the beat  frequency
mechanism, the sonic-point beat frequency model, is due to Miller et al.
(1998). The model faced  criticism in the  work of M\'endez et al.
(1998) when the $\Delta \nu$ was found to
be less than half the  burst oscillation frequency $\nu_b$. This could
be explained in the sonic-point  model (Lamb \& Miller 2001) by taking into
account the inward plasma velocity (see JMK for details).  However, in the
recent observations  of JMK, the  $\Delta \nu$ is  significantly larger
than half of $\nu_b$ causing  further problems. The only change in the
flow  pattern previously  described  that could  produce the  observed
change in  $\Delta \nu$  is that in  which the accretion  disk changes
from prograde  to retrograde when  the lower kHz  QPO is seen  to move
from ~ 750 to 800 Hz.

In contrast to  the beat frequency, the sonic-point  and other
models, Osherovich \& Titarchuk  (1999) and Titarchuk  \&
Osherovich  (1999) take the  inner  QPO to  be the Keplerian
accretion  disk, leading  to a substantially  smaller radius for
the compact  object that is hard to explain from  any of the known
neutron star models. Li et al. (1999) showed that the observed
compact star 4U~1728$-$34 can be fitted to  a strange star model
that uses the realistic strange matter  equation of state of Dey
et al. (1998). The higher QPO in the Osherovich \& Titarchuk
(1999) model is due the modification of Keplerian frequency under
the influence of the Coriolis force in a rotating frame of
reference.

The present letter deals with a  model which takes the lower QPO
to be the  Keplerian motion of  the accreting  fluid around
compact strange star, rotating very  fast. It also seeks to
explain  the higher QPO as being due to viscous effects  that
lead a shock formation (Chakrabarti 1989, 1996;  Molteni et al.
1996b,  hereinafter MSC) and  solves the
hydrodynamic  equations  in the  presence  of a  {\it
pseudo-Newtonian potential} (Mukhopadhyay 2002a; Mukhopadhyay \&
Misra 2003, hereinafter MM) which  describes the relativistic  properties of
accretion disk close to the compact object  with an effective
cutoff for an appropriate boundary  condition  that  needs  the
accreting particles  to  go  to asymptotic zero velocity. The
same  model applied to black holes would allow them to get to
luminal velocities.

The formation  of shock  in accretion disk  around compact object
was discussed  by several  independent  groups (e.g. Sponholz \&
Molteni 1994; Nabuta  \& Hanawa 1994; Molteni et al.
1994; Yang \&  Kafatos 1995; Chakrabarti  1996;
Molteni et al. 1996a;  MSC; Lu \&  Yuan 1997;
Chakrabarti  \& Sahu  1997) either  by analytical approach or
numerical simulations.  Recently, Mukhopadhyay (2002b) showed,
even two shocks are possible to form in accretion disk around
neutron star (or other  compact object which has hard surface).
It was shown in MSC that  the shock location may oscillate in the
disk and this  oscillatory nature  is directly related  to the
cooling and advective time  scale of  the matter. Subsequently it
was  shown that corresponding oscillation frequency is related to
the location of the shock and  the observed  QPO frequency for   the  various
black   hole candidates  could   be explained.  They (MSC)
showed  the theoretical  calculations for  QPO tally with that of
observation for black hole candidate (which is of the order of
Hz), say  GS 339-4 and  GS 1124-68. As the shock  location comes closer to the
compact  object, the  QPO frequency increases (MSC).  In case  of
accretion flow around compact object other than black hole, the
shock(s) may form even closer to  the compact object with respect
to the case of black hole (Mukhopadhyay  2000b).  Thus the
corresponding QPO frequencies are expected to  be higher as of
kilohertz order which  could explain the origin of  observed
higher frequency (HF) oscillations of kilohertz QPO for compact
objects.

Recently, MM prescribed a couple  of  potentials  to  describe  the  time
varying  relativistic properties of accretion disk  around
compact objects.  They considered the Keplerian accretion disk
and proposed modified gravitational force which  could give  the
Keplerian  angular frequency  of  the accreting matter around the
compact object. As mentioned above, the lower frequency (LF)
oscillation of the kilohertz  QPO may arise due to the
Keplerian motion of the accreting matter  towards compact object,
and if we know the radius of  the Keplerian orbit, following MM
we can calculate the LF.

In  this letter  we  will calculate  the  pair of  kHz  QPOs for
fast rotating  candidates  mainly  for  4U  1636$-$53
($\nu=582$Hz)  and  KS 1731$-$260  ($\nu=523.92$Hz). To  calculate
HF,  we will  more  or less follow MSC but  with the necessary
modification using  the model of MM due to the rotation of
compact object  (the work of MSC was confined with non-rotating
black holes  only). For LF, we will  follow MM where the {\it
pseudo-Newtonian  potentials and corresponding  forces} are given
to  describe the  Keplerian motion  of the  accreting matter.  In
next section, we will  briefly describe the basic viscous  set of
accretion disk equation and formalism to  calculate QPO
frequencies.  Then in \S 3, we will  tabulate QPO frequencies
from our  theory and compare with observation and finally in \S
4, we will present a summary.

\section{Basic Equations and formalism }

We  will follow  Chakrabarti (1996)  to  describe the  viscous set  of
equations for accretion disk which are given as
   \begin{eqnarray}
   \nonumber
&&   -4\pi x\Sigma v=\dot{M},\hskip0.5cm
   v\frac{dv}{dx}+\frac{1}{\rho}\frac{dP}{dx}-\frac{\lambda^2}{x^3}+F(x)=0, \\
   \nonumber
&&   v\frac{d\lambda}{dx}=\frac{1}{\Sigma x}\frac{d}{dx}\left[x^2\alpha\left
(\frac{I_{n+1}}{I_n}P+v^2\rho\right)h(x)\right], \\
   \nonumber
&&   \Sigma vT\frac{ds}{dx}=\frac{vh(x)}{\Gamma_3-1}\left(\frac{dP}{dx}-
\Gamma_1\frac{p}{\rho}\frac{d\rho}{dx}\right)=Q^+-Q^-. \\
   \label{diskeq}
   \end{eqnarray}

Here, throughout our calculations  we express the radial
coordinate in unit of  $GM/c^2$, where $M$ is mass  of the
compact star,  $G$ is the gravitational constant and $c$ is  the
speed of light. We also express the velocity  in unit of  speed
of light  and the angular  momentum in unit   of  $GM/c$.
Following   Cox  \&   Giuli  (1968),   we  define
$\Gamma_1,\Gamma_3$  and  to  define  vertically  integrated
density, $\Sigma$ and pressure, $W$ follow Matsumoto et al.
(1984). $\beta$ and $h(x)$ are defined as the ratio  of gas
pressure to total pressure and half-thickness    of     the
disk,    $h(x)=c_s    x^{1/2}F^{-1/2}$ respectively.  We follow
Mukhopadhyay (2002a)  and MM  to  define the effective
gravitational  pseudo-Newtonian   force,  $F(x)$. We consider the
adiabatic flow, where  the  speed of sound is given as 
$c_s^2=\frac{\gamma  P}{\rho}$, related to the temperature ($T$) of the system as,
$c_s=\sqrt{\frac{\gamma kT}{m_p}}$. We consider the magnetic field strength 
in the disk to be negligible compared to the viscous effect 
and the heat evolved is solely due to
viscosity which is defined as $Q^+=\frac{W_{x\phi}^2}{\eta}$ where  
$W_{x\phi}$  and  $\eta$ are  the viscous  stress  tensor  and coefficient of  viscosity
respectively. For simplicity, the heat lost is considered
proportional to the heat gained by the flow. Thus the net heat
contained in the flow is chosen to be $Q^+f$, where $f$ is the
cooling factor, which is close to $0$ and $1$ for cooling and
heating dominated flow respectively. $\alpha$ is Shakura-Sunyaev
(Shakura \& Sunyaev  1973) viscosity parameter, $s$ is entropy
density. To get the thermodynamic properties of disk, we need to solve (\ref{diskeq}).
  
It is  shown in MSC that when  shock is formed in  accretion disk, the
cooling ($t_{\rm cool}$) and advection ($t_{\rm adv}$) time scale of matter from
the shock location  to the inner edge of the  disk are responsible for
the oscillatory  behaviour of shock that  is related to  the QPO. When
the  physical condition  of the  disk is  such that  $t_{\rm cool}\sim
t_{\rm  adv}$, corresponding  QPO can  be found  and  that oscillation
frequency  is of  the  order  of $1/t_{\rm  adv}$.  Therefore, we  can
calculate
\begin{eqnarray}
t_{\rm adv}=\int_{x_s}^{x_{\rm in}}\frac{dx}{v}
\label{ts}
\end{eqnarray}
where, $x_s$ and $x_{\rm in}$ are the location of shock and inner edge
of  the disk respectively.   In our  case, $x_{\rm  in}$ is  the outer
radius of the compact object.

Therefore, to calculate the QPO frequency, the
detailed knowledge of the thermodynamic character of the disk from
shock to the surface of the compact object is very
essential. Once this is known for a particular  accretion flow
around compact  object,  the corresponding QPO frequency can  be
calculated  which is basically  HF.  Mukhopadhyay (2002b)
already indicated that the kilohertz QPO  can be calculated from
the accretion disk model around slowly rotating neutron star.
Here we would like to consider the rapidly rotating compact star  which
results the shift  in shock location with respect to that  of
non-rotating cases and calculate the QPO frequency for different
viscosity.

MM proposed  a couple of  pseudo-Newtonian potentials to
describe the time  varying properties of  the Keplerian
accretion disk. On the other hand, LF may arise due to
the  Keplerian motion of the accreting fluid. One of  the
potentials  proposed by  MM, that  can describe  the Keplerian
angular frequency of the accretion flow is given as
\begin{eqnarray}
2\pi\nu_K=\Omega_K=\frac{1}{x^{3/2}}\left[1-\left(\frac{x_{\rm ms}}{x}\right)+
\left(\frac{x_{\rm ms}}{x}\right)^2\right]^{1/2}.
\label{lbo}
\end{eqnarray}
Here $x_{\rm ms}$  is the radius of marginally  stable orbit, for Kerr
geometry that was given by Bardeen (1973) as
\begin{eqnarray}
\nonumber
x_{\rm ms}  &=&  3 + Z_2 \mp[(3-Z_1)(3+Z_1+2Z_2)]^{1/2} \\
\nonumber
Z_1 & =&  1 +(1-J^2)^{1/3}[(1+J)^{1/3}+(1-J)^{1/3}],\\
\nonumber
Z_2 & = & (3J^2+Z_1^2)^{1/2},
\end{eqnarray}
where  the '-' ('+')  sign is  for the  co-rotating
(counter-rotating) flow and $J$ (which varies from $0-1$)  is the
specific angular momentum of compact object.  In our
present  work, we propose $\nu_K$ as the lower kilohertz
frequency  (LF) that varies  with the angular  frequency of the
compact  object. If we know  the angular frequency  of the compact
object and  the Keplerian radius of the  corresponding accretion
disk, LF can be calculated for that particular candidate.

\section{ QPO frequencies}

The observed range  of frequency of the QPOs  for different
candidates are quite large. However, there  exists a relation
between the LF and the  HF which  relates them  to  the
intrinsic  characteristic of  the central compact object like the
mass-radius relation, spin, etc. Also it has been seen that for a
particular candidate, there is a variability in  the frequency
range for  different times  of  observation, however keeping  the
co-relation  between the  HF  and the  LF constant.  We encash
upon these to our present interpretation.



\vbox{
\vskip 0.0cm
\hskip 1.7cm
\centerline{
\psfig{figure=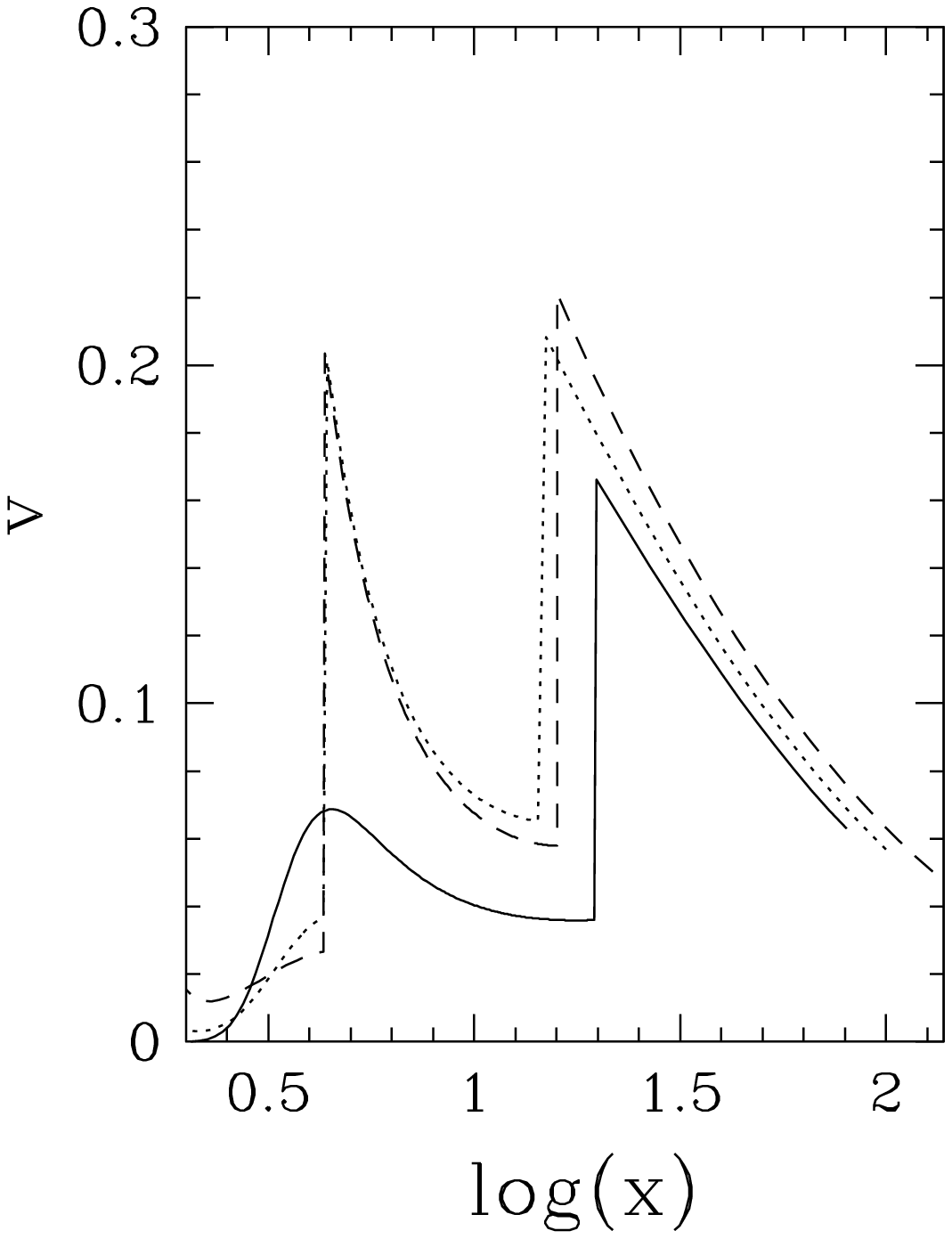,height=14.0truecm,width=14.0truecm}}}
\vspace{-6.2cm}
\noindent {\small {\bf Fig. 1}:
Variation of accretion speed in unit of light speed as a function of
radial coordinate around 4U 1636-53. Solid, dotted and dashed curves are respectively for
 (i) $\alpha=0, f=0$, (ii) $\alpha=0.02, f=0.2$, (iii) $\alpha=0.05,
f=0.5$.
Other parameters are $J=0.2877, \lambda_c=3, \dot{M}=1, \beta=0.03$.
}
\vskip0.2cm

\begin{table*}[htbp]
\center{\large Table-1}
\begin{center}
\hskip-2.0cm
\begin{small}
\begin{tabular}{|c|c|c|c|c|c|c|c|c|c|c|c|}
\hline
\hline
 source & $\alpha$ & f & $r_s$ & $r_k$ & J & M  & R & $\nu_h$ & $\nu_K$ &
\multicolumn{2}{c|}{observed} \\
\cline{11-12}
& & & (km) & (km) & & $(M_\odot$) & (km) & Hz & Hz & HF(Hz) & LF(Hz) \\
\hline
  & 0.05 & 0.5 & 15.37 & 18 & 0.2877 & 1.18 & 7.114 & 1030.8 & 719.7 & & \\
\cline{2-10}
4U~1636$-$53  &0.02 & 0.2 & 14.27 & 19 & 0.2877 & 1.32 & 7.23 & 1019.2 & 705.2 & 1030 & 700 \\
\cline{2-10}
& 0 & 0 & 13 & 17 & 0.2877 & 0.991 & 6.828 & 1005.2 & 715.2 & &\\
\hline
 & 0.05 & 0.5 & 15.37 & 15.2 & 0.2585 & 1.106 & 7.013 & 1155.4 & 907.6 & & \\
\cline{2-10}
KS~1731$-$260  &0.02 & 0.2 & 14.27 & 16 & 0.2585 & 1.23 & 7.16 & 1174 & 892.6 & 1159 & 898 \\
\cline{2-10}
& 0 & 0 & 13 & 14.5 & 0.2585 & 0.893 & 6.64 & 1151.9 & 863.1 & &\\
\hline
\hline
\end{tabular}
\end{small}
\end{center}
\end{table*}

Our present  consideration will mainly be focussed upon the
QPOs of 4U~1636$-$53  and  KS~1731$-$260 which  are  among  the
fast  rotating compact stars  having angular frequency $\sim
582$Hz ($J=0.2877$) and $\sim 524$Hz ($J=0.2585$) respectively.
Both these stars have displayed long lasting superbursts. Long
lasting superbursts find a natural explanation in the diquark
surface pairing after these pairs are broken (Sinha et al. 2002).
Thus, it is justified to invoke the realistic strange star
model for these stars and our inputs, namely the star mass and
radius is taken from this model of Dey et al. (1998). We estimate
the LF from the stellar frequency and the Keplerian radius  of the
disk. As HF is directly related to the accreting matter speed, in
Fig. 1, we show the variation of accretion speed, for various
viscosities of the accretion disk,  around 4U~1636$-$53. As the
angular frequency of KS~1731$-$260 is close to 4U~1636$-$53,
profile of matter speed around KS~1731$-$260 does not differ
significantly from that around 4U~1636$-$53   and   is not shown.
Obviously, we consider those cases where shock does form in the
accretion disk which is responsible for HF. In case of inviscid
flow, there is only a single shock  formation in the accretion
disk, while for the viscous flow, shock is formed twice, the
shock locations being listed  in Table-1. Now following
(\ref{ts})  and (\ref{lbo}) we can calculate HF and LF for
various physical parameters of the disk and compact objects.
The inner boundary condition of the model decelerates the speed of the 
infalling matter close to the stellar surface, and hence loses its energy. 
This in turn reduces the temperature of the falling matter to a considerable 
amount and hence the contribution of the energy transfer to the surface
burst phenomena (Kuulkers 2002) is negligible.

In the same table, it is  shown that the values of `$\alpha$' when
varied  from 0 to  0.05 reproduce an  admissible value of  the QPO
frequencies.  At  this point,  it  is  necessary  to mention that
the mass-radius relation of the central  compact source also
plays a vital role  in the determination  of the frequencies.
Compact  objects with hard surface  are generally considered  as
neutron stars, but  in our study we find that the mass-radius
relation obtained from the neutron star  equations of  state
(EOS)  are  not  satisfactory  enough  to reproduce the matching
QPOs. They need to be more compact and hence we had to look  for
an  alternative to  the  conventional neutron  star EOS. As
already hinted, recent developments in  the theory of  strange
stars and their EOS gave us the alternative choice. We choose the
EOS (SS1 as in Li et al. 1999) for  compact strange  star
adapted from the model of Dey  et al. (1998).  In Table-1, we
enlist the values of QPO frequencies for both the candidates
4U~1636$-$53 and KS~1731$-$260 for different cases of variation
of the parameters. The corresponding chosen mass and radius of
the star are  all in the `allowed range' as per the model  of Dey
et  al. (1998).  The $\nu_h$s  are calculated  from fluid
dynamical model  and compared with  observed HF.  Also, $\nu_K$s
are calculated theoretically  and compared with observed LF. It
is very exciting to see that our calculated $\nu_h$ and $\nu_K$
are very close to the observed HF and LF respectively. The
results not only reflect the success of the model, but also
reinforces the claim of the existence of strange stars.

Considering the stellar spin frequency is half of the assumed values
(Titarchuk 2003) the change of $\nu_K$ is negligible ($<<1\%$) (eq.
[\ref{lbo}]). In this situation that matter will adjust itself in the disk at the 
cost of changing other physical parameters, namely, angular momentum ($\lambda$), 
sonic points etc. (Mukhopadhyay 2003) to keep unchanged
the shock location as well as $\nu_h$, at least for $|J|\leq 0.3$.
Similarly, with the presence of moderate magnetic field the shock can be formed
at a same location for a different choices of $\lambda$, $\alpha$, $s$ etc.

\section{Summary}
It  is beyond  doubt that  the  origin of  the kilohertz  QPOs
is  the accretion disk centering  around a compact object.  We
model here, the source of origin of the kilohertz QPOs in the
light of fluid dynamical calculation of the inflowing matter in the
accretion disk. It is perhaps for the  first time such a theoretical
calculation is presented which  matches with the results of the
observation to a very satisfactory  limit. Although we have used
here two of  the fastest rotating  candidates, 4U~1636$-$53 and
KS~1731$-$260, to compare our  results and also identify  them as
strange stars rather than conventional neutron stars, a detailed
study of  all the other candidates in this  model is necessary.
The choice of the two above mentioned stars was also motivated by
the fact that they have recently displayed very long lasting
superbursts (Sinha et al. 2002).

Out of the QPO pair, higher one is directly related to the 
accretion flow around the compact object. If shock is formed in the 
accretion disk and the advective time-scale is of the same order as the cooling 
time-scale in the disk from the shock location to the stellar 
surface, we immediately expect higher kHz QPO. In earlier works, 
theoretical calculations based on this idea, had been carried out for 
the black hole candidates GS~339$-$4 and GS~1124$-$68 (MSC). In the present letter,
similar procedure is applied for other stellar candidates and 
with a modification for inclusion of rotation of the compact objects.
The results ended up in the discovery of those candidates as strange stars.
Lower QPO frequency arises, when we consider the 
Keplerian motion of the accretion fluid.
The most important aspect of our work is that we have not made use of  any
`toy-model' to describe the QPOs, but they arrived naturally from
the hydrodynamical calculations.

\begin{acknowledgements}
BM acknowledges a discussion made three years back with Sandip K. Chakrabarti. 
\end{acknowledgements}


{}

\end{document}